\documentclass[10pt,pra,twocolumn,showpacs,floatfix]{revtex4}
\usepackage{amssymb}
\usepackage{graphicx}
\usepackage{times}
\usepackage{amsmath}

\begin{document}

\title{Toward demonstrating controlled-X operation based on continuous
variable four-partite cluster state and quantum teleporters}
\author{Yu Wang}
\author{Xiaolong Su}
\email{suxl@sxu.edu.cn}
\author{Heng Shen}
\author{Aihong Tan}
\author{Changde Xie}
\author{Kunchi Peng}
\affiliation{State Key Laboratory of Quantum Optics and Quantum Optics Devices, Institute
of Opto-Electronics, Shanxi University, Taiyuan, 030006, People's Republic
of China}

\begin{abstract}
One-way quantum computation based on measurement and multipartite
cluster entanglement offers the ability to perform a variety of
unitary operations only through different choices of measurement
bases. Here we present an experimental study toward demonstrating
the controlled-X operation, a two-mode gate, in which continuous
variable (CV) four-partite cluster states of optical modes are
utilized. Two quantum teleportation elements are used for achieving
the gate operation of the quantum state transformation from input
target and control states to output states. By means of the optical
cluster state prepared off-line, the homodyne detection and
electronic feeding forward, the information carried by the input
control state is transformed to the output target state. The
presented scheme of the controlled-X operation based on
teleportation can be implemented nonlocally and deterministically.
The distortion of the quantum information resulting from the
imperfect cluster entanglement is estimated with the fidelity.
\end{abstract}

\maketitle

\section{Introduction}

Developing faster and faster computers is a long process. Quantum computers
(QCs) based on the fundamental principles of quantum physics, such as
coherent superposition and entanglement of quantum states, promise
super-fast and powerful computation capacities for the future over present
classical computers. In recent years, esoteric and attractive ideas about
quantum computers were being converted into visible realization step by step
along with experimental demonstrations of various quantum logic operations
in both discrete variable (DV) and continuous variable (CV) domains \cite%
{Monroe1995,Pittman2002,Kaler2003,Leibfried2003,Li2003,Zhao2005,Yoshikawa2008,Miwa2009}%
. Different models for quantum computation, typically the conventional
circuit model and the so-called cluster model, were proposed and
experimentally realized \cite%
{Nielsen2000,Chuang1998,Gottesman1999,Gulde2003,Lu2007,Politi2009,Raussendorf2001,Walther2005,Prevedel2007,Chen2007}%
. According to the cluster-state model initially proposed by Raussendorf and
Briegel \cite{Raussendorf2001},the actual computation can be completed only
by a sequence of single-qubit projective measurements with classical
feed-forward of the measured results with the help of the special
multiparticle entanglement between individual subsystems of the cluster
state prepared off-line. Due to the irreversibility of measurements the
cluster-based QC is also inherently time-irreversible, and thus is named the
one-way QC. The most important feature of cluster QCs is its universality,
i.e. any quantum circuit can be implemented on a suitable cluster state \cite%
{Raussendorf2001}. Exploiting four-photon cluster states, the performances
of one-way DV QC were experimentally demonstrated \cite%
{Walther2005,Prevedel2007,Chen2007}.

In 1999, Lloyd and Braunstein extended the quantum computation to the CV
region and derived the necessary and sufficient conditions for achieving
universal CV QC \cite{Lloyd1999}. Successively, the CV QC with optical
coherent states \cite{Ralph2003} and the encoding schemes for CV computation
were proposed \cite{Gottesman2001}. The scalable CV error correction
routines used for CV QCs were theoretically investigated \cite%
{Braunstein1998} and experimentally demonstrated with CV multipartite
entanglement of optical modes \cite{Aoki2009}. Based on off-line squeezed
states and quantum nondemolition (QND) interaction between the quadrature
components of two optical modes, Furusawa's group realized the QND sum gate
\cite{Yoshikawa2008} in the CV QC circuit model proposed by Filip et al.
\cite{Filip2005}. Very recently, the same group achieved an experimental
demonstration of the principles of a universal one-way quantum quadratic
phase gate over CVs, in which a two-mode cluster state of light is involved
\cite{Miwa2009}. In contrast to the generation systems of DV cluster states
of single-photons, CV cluster states of optical modes can be prepared
unconditionally and quantum computations with CV cluster can be performed
deterministically \cite{Zhang2006,vanLoock2007}. Following the first
theoretical proposal on universal QC with CV cluster states \cite%
{Menicucci2006}, various protocols of cluster CV QCs for experimental
implementation are studied in detail \cite{vanLoockJOSA2007,Tan2009,Gu2009}.
In experiments, CV cluster states involving four optical modes were
successfully prepared \cite{Su2007,Yukawa2008}.

It was pointed out that the single-qubit and two-qubit gates used in
teleportation are sufficient to construct even the most complex QCs \cite%
{Lloyd1992}. Generally, in QCs quantum information is propagated via
teleportation networks, thus quantum teleporters are key elements of
building QCs \cite{Hemmer2009}. By the end of the last century quantum
teleportation was been experimentally realized with both DV \cite%
{Bouwmeester1997,Boschi1998,Riebe2004,Barrett2004,Huang2004,Sherson2006,Olmschenk2009}
and CV \cite{Furusawa1998,Bowen2003,Zhang2003} protocols. Later, the
teleportation networks were achieved with single-photon multiparticle
polarization-entanglement \cite{Zhao2004} and multipartite
quadrature-entanglement of optical modes \cite{Yonezawa2004}, respectively.
These successful experiments on teleportation provide the fundamental
technology to construct QCs.

In this article, we present an experimental study toward demonstrating the
controlled-X operation, which is an analog of a two-qubit contolled-NOT gate
in the CV regime \cite{Menicucci2006}. This controlled-X operation (it is
also called the sum gate) based on utilizing CV four-partite cluster states
of optical modes is different from that achieved in Refs. \cite%
{Yoshikawa2008} and \cite{Miwa2009}. The QND sum gate in Ref. \cite%
{Yoshikawa2008} belongs to the typical circuit model and the quadratic phase
gate in Ref. \cite{Miwa2009} only achieves a single-mode operation involving
a two-mode cluster state. According to our previous theoretical design (see
the Sec. V in Ref. \cite{Tan2009}) we experimentally explored the scheme of
implementing the one-way controlled-X operation, a two-mode gate using the
linear four-partite cluster state of optical modes, homodyne detectors and
electronic feeding forward. In the operation system two CV quantum
teleportation elements are included, which are used for teleporting the
information of the input target and control states to the output states. The
experimental results show that the amplitude quadrature of the output target
is displaced a certain amount set by the input control state according to
the requirement of the controlled-X operation. The measured variances of the
quadratures of the output states in the case exploiting cluster quantum
resources are about 1.8 dB below the shot noise limit (SNL) determined by
the vacuum noise levels of coherent states without the existence of cluster
entanglement. Based on the nonlocal and deterministic entanglement feature
of CV cluster states the presented controlled-X operation can be also
implemented nonlocally and deterministically. The construction of the
experimental system for the controlled-X operation exhibits the key role of
quantum teleportation in QCs, obviously. The fidelities of the output target
and control states are calculated, respectively, and both of them surpass
their classical limit. Since the squeezing level of the resources of the
cluster state is not high enough (only $\sim $ 3 dB) the entanglement
between output target and control modes, which is necessary for further
quantum information processing, was not observed in the present experiment.
However, we provide a scalable experimental system and scheme toward
achieving the controlled-X operation.

\section{Operation principle}

The Hamiltonian of the controlled-X operation can be written as $\hat{H}=-%
\hat{X}_{c}\hat{Y}_{t}$, where $\hat{X}=\hat{a}+\hat{a}^{+}$ and $\hat{Y}=(%
\hat{a}-\hat{a}^{+})/i$ are the amplitude and phase quadratures of an input
optical mode $\hat{a}$, and the subscripts $c$ and $t$ denote the control
and target modes, respectively. The ideal input-output relation of the
controlled-X operation are given by \cite{Lloyd1999},%
\begin{eqnarray}
\hat{X}_{c}^{out} &=&\hat{X}_{c},\qquad \hat{X}_{t}^{out}=\hat{X}_{t}-\hat{X}%
_{c}, \\
\hat{Y}_{c}^{out} &=&\hat{Y}_{c}+\hat{Y}_{t},\qquad \hat{Y}_{t}^{out}=\hat{Y}%
_{t}.  \notag
\end{eqnarray}%
Through the controlled-X operation the input control (target) variable $\hat{%
X}_{c}$ $(\hat{Y}_{t})$ is added to the output target (control) variable $%
\hat{X}_{t}^{out}$ $(\hat{Y}_{c}^{out})$, while $\hat{X}_{c}$ $(\hat{Y}_{t})$
remains unchanged. The operation results in a phase-space displacement on
the amplitude quadrature of the target by an amount determined by the
position (amplitude) eigenvalue $\hat{X}_{c}$ of the control and possibly
establishes the quantum entanglement between the output target and control
modes, which is just the aim of the controlled-X gate.
\begin{figure}[tbp]
\begin{center}
\includegraphics[width=8cm,height=7cm]{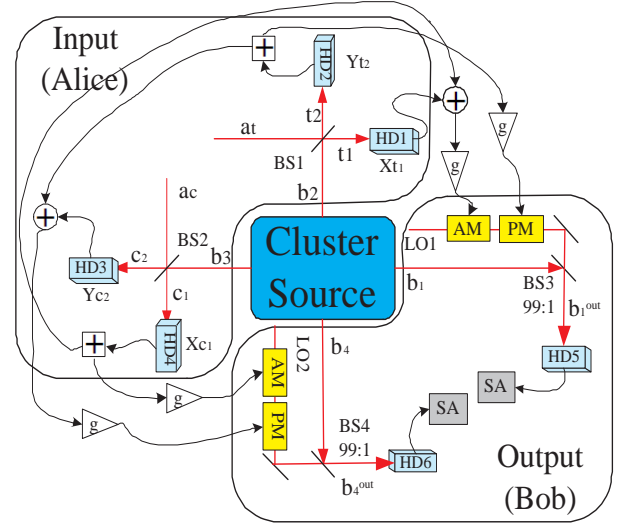}
\end{center}
\caption{(Color online) Schematic of controlled-X operation. b1-4:
four submodes of CV linear four-partite cluster state from the
cluster source; HD1-6: homodyne detection system; AM: amplitude
modulator; PM: phase modulator; BS1 and 2: 50\% beamsplitter; BS3
and 4: 99:1 beamsplitter; at: input target mode; ac: input control
mode; c1 and 2: the output field from BS2; t1 and 2: the output
field from BS1; g: gain of feed forward circuit; $\boxplus $: power
splitter; $\oplus $: positive power combiner; LO: auxiliary local
oscillation beam; and SA: spectrum analyzer.}
\end{figure}

\section{Experimental method}

Figure 1 shows the schematic of the controlled-X operation with the
four-partite linear CV cluster state of the optical field. The four submodes
$\hat{b}_{1}-\hat{b}_{4}$ of the cluster state are generated by linearly
combining four quadrature-squeezed states of light produced from a pair of
nondegenerate optical parametric amplifiers (NOPAs) operating at
deamplification and below the oscillation threshold \cite{Su2007}. Compared
to our previous cluster generation system \cite{Su2007} only a slight change
is made, that is, a $1:1$ beamsplitter (BS1) in the old system is replaced
by a $4:1$ beamsplitter thus eliminating the effect of the antisqueezing
component on the squeezing component to the largest extent \cite%
{Tan2009,Yukawa2008}. The squeezed correlation variances of the cluster
state are expressed by \cite{Tan2009}
\begin{eqnarray}
\hat{Y}_{b_{1}}-\hat{Y}_{b_{2}} &=&\sqrt{2}e^{-r}\hat{Y}_{a_{1}}^{(0)}, \\
\hat{X}_{b_{1}}+\hat{X}_{b_{2}}+\hat{X}_{b_{3}} &=&\frac{\sqrt{10}}{2}e^{-r}%
\hat{X}_{a_{2}}^{(0)}-\frac{\sqrt{2}}{2}e^{-r}\hat{Y}_{a_{4}}^{(0)}, \\
-\hat{Y}_{b_{2}}+\hat{Y}_{b_{3}}+\hat{Y}_{b_{4}} &=&-\frac{\sqrt{10}}{2}%
e^{-r}\hat{X}_{a_{3}}^{(0)}+\frac{\sqrt{2}}{2}e^{-r}\hat{Y}_{a_{1}}^{(0)}, \\
\hat{X}_{b_{3}}-\hat{X}_{b_{4}} &=&-\sqrt{2}e^{-r}\hat{Y}_{a_{4}}^{(0)}.
\end{eqnarray}%
where $r$ stands for the squeezing parameter of the squeezed states, which
depends on the strength and the time of parametric interaction in NOPA. We
assumed that $r$ of the four squeezed states is identical, which is not
difficult to reach experimentally by balancing the configuration and system
parameter of NOPAs. The values of $r$ are between zero and positive infinite
with $r=0$ for no squeezing and $r\rightarrow \infty $ for the ideal
squeezing. However, the ideal squeezing limit can not be achieved in
experiments since it requires infinite energy. $\hat{X}_{ai}^{(0)}$ and $%
\hat{Y}_{ai}^{(0)}$ ($i=1-4$) denote the quadrature amplitudes and phases of
seed optical beams initially injected into NOPAs, respectively \cite%
{Tan2009,Su2007}. The variances of the seed beams in coherent states are
normalized, that is\textbf{\ }$V(\hat{X}_{ai}^{(0)})=V(\hat{Y}_{ai}^{(0)})=1$%
.\ The submodes $\hat{b}_{2}$, $\hat{b}_{3}$ and $\hat{b}_{1}$, $\hat{b}_{4}$
are distributed to input (Alice) and output (Bob) of the controlled-X gate,
respectively.

At Alice, submodes $b_{2}$ and $b_{3}$ are mixed with the target ($\hat{a}%
_{t}$) and the control ($\hat{a}_{c}$) signals on $1:1$ beamsplitters BS1
and BS2, respectively. The amplitude and phase quadratures of the output
target (control) modes $\hat{t}_{1}$ and $\hat{t}_{2}$ ($\hat{c}_{1}$ and $%
\hat{c}_{2}$) from BS1 (BS2) are detected by a pair of balanced homodyne
detectors HD1 and HD2 (HD3 and HD4) actively locked to be $0$ and $90%
{{}^\circ}%
$ out of phase, respectively. The measured conjugate quadratures of optical
modes are denoted by $\hat{X}_{t1}$, $\hat{Y}_{t2}$ and $\hat{X}_{c1}$, $%
\hat{Y}_{c2},$ respectively, for the target and the control. The measured
photocurrents of $\hat{X}_{c1}$\ and $\hat{Y}_{t2}$\ are split by two power
splitters ($\boxplus $), respectively. Then a half of the photocurrent $\hat{%
X}_{c1}$ ($\hat{Y}_{t2}$)\ is added to $\hat{X}_{t1}$\ ($\hat{Y}_{c2}$)\ by
a positive power combiner ($\oplus $). The photocurrents ($\hat{X}_{t1}+\hat{%
X}_{c1}$) and ($\hat{Y}_{t2}+\hat{Y}_{c2}$), as well as the remaining other
half of $\hat{X}_{c1}$ and $\hat{Y}_{t2}$\ are sent to Bob through four
classical channels with a suitable electronic gain $g$ ($\lhd $), where Bob
uses them to impose amplitude and phase modulations on two bright laser
beams in coherent states (LO1 and LO2) by means of amplitude (AM) and phase
(PM) modulators, respectively. By mixing the modulated LO1 (LO2) with the
submode $\hat{b}_{1}$ ($\hat{b}_{4}$) remaining by Bob on a $99:1$ highly
reflective beamsplitter BS3 (BS4), a displacement in the phase-space
proportional to the amplitude and phase modulations is imposed on $\hat{b}%
_{1}$ ($\hat{b}_{4}$). Because of cluster entanglement among the four
submodes $\hat{b}_{1}-\hat{b}_{4}$, Alice's two Bell-state detections
collapse both submodes $\hat{b}_{1}$ and $\hat{b}_{4}$ into a state
conditioned on the measurement outcomes $\hat{X}_{t1}$, $\hat{Y}_{t2}$ and $%
\hat{X}_{c1}$, $\hat{Y}_{c2}$. The amplitude quadrature $\hat{X}_{t}^{out}$ (%
$\hat{X}_{c}^{out}$) and the phase quadrature $\hat{Y}_{t}^{out}$ ($\hat{Y}%
_{c}^{out}$) of the displaced outcome $\hat{b}_{1}^{out}$ ($\hat{b}_{4}^{out}
$) are expressed by \cite{Tan2009}%
\begin{eqnarray}
\hat{X}_{t}^{out} &=&\hat{X}_{b_{1}}+\sqrt{2}g\hat{X}_{t_{1}}+\sqrt{2}g\hat{X%
}_{c_{1}}  \notag \\
&=&\sqrt{\frac{5}{2}}e^{-r}\hat{X}_{a_{2}}^{(0)}-\sqrt{\frac{1}{2}}e^{-r}%
\hat{Y}_{a_{4}}^{(0)}+\hat{X}_{t}-\hat{X}_{c}, \\
\hat{Y}_{t}^{out} &=&\hat{Y}_{b_{1}}-\sqrt{2}g\hat{Y}_{t_{2}}  \notag \\
&=&\sqrt{2}e^{-r}\hat{Y}_{a_{1}}^{(0)}+\hat{Y}_{t}, \\
\hat{X}_{c}^{out} &=&\hat{X}_{b_{4}}-\sqrt{2}g\hat{X}_{c_{1}}  \notag \\
&=&\sqrt{2}e^{-r}\hat{Y}_{a_{4}}^{(0)}+\hat{X}_{c}, \\
\hat{Y}_{c}^{out} &=&\hat{Y}_{b_{4}}-\sqrt{2}g\hat{Y}_{t_{2}}+\sqrt{2}g\hat{Y%
}_{c_{2}}  \notag \\
&=&-\sqrt{\frac{5}{2}}e^{-r}\hat{X}_{a_{3}}^{(0)}+\sqrt{\frac{1}{2}}e^{-r}%
\hat{Y}_{a_{1}}^{(0)}+\hat{Y}_{t}+\hat{Y}_{c}.
\end{eqnarray}

In cluster QC language the displacement operation is equivalent to
feed-forward the measurement results of the submodes $\hat{b}_{2}$ and $\hat{%
b}_{3}$ on the remained submodes $\hat{b}_{1}$ and $\hat{b}_{4}.$ In the
experiment, the gain $g$ of all feed-forward circuits is carefully adjusted
to 1, which corresponds to realizing a controlled-X operation in Eq.(1). It
is obvious that from Eqs. $(6)-(9)$ that in the limit of infinite squeezing (%
$r\rightarrow \infty $), the amplitude quadrature of the output target, $%
\hat{X}_{t}^{out}$, was displaced an amount $\hat{X}_{c}$ determined by the
amplitude of the input control field. It means that the ideal controlled-X
operation was completed. For practical experiments with a finite value of $r$%
, some noise resulting from imperfect squeezing will be added on the $\hat{X}%
_{t}^{out}$, thus the fidelity of the outcome state will reduce. However,
the fidelity using finite cluster entanglement of $r\neq 0$ will be higher
than its classical limit which is measured in the case without the existence
of cluster state. To verify the performance of the controlled-X operation
the outcome values of $\hat{X}_{t}^{out}$, $\hat{Y}_{t}^{out}$ and $\hat{X}%
_{c}^{out}$, $\hat{Y}_{c}^{out}$ are detected by the homodyne detectors, HD5
and HD6, respectively.

The cluster source comprises a master-pump laser, a pair of NOPAs and some
linearly optical elements, which are not shown in Fig. 1 (see Ref. \cite%
{Su2007} for details). A continuous wave intracavity frequency-doubled and
frequency-stabilized Nd:YAP/KTP (Nd-doped YAlO3 perovskite / potassium
titanyl phosphate) laser (made by Yuguang Co. Ltd., F-VIB) \cite%
{Li2002,Jia2004} serves as the master laser of the experimental system. The
output second-harmonic wave at 540 nm is used for the pump laser of the two
NOPAs to produce four two-mode squeezed states at 1080 nm via intracavity
frequency down-conversion. The four squeezed states are transformed to a
linear four-partite cluster state with optical beamsplitters \cite{Su2007}.
In the system, all the optical beams at coherent states, including the
target signal ($\hat{a}_{t}$), the control signal ($\hat{a}_{c}$), LO1, LO2,
and the local oscillation beams used in homodyne detectors (HD1-6),
originate from the subharmonic output of the master laser at 1080 nm. During
the experiment the pump power of the NOPAs are kept at 175 mW which is below
the oscillation threshold of 230 mW, and the intensity of the signals into
NOPAs at 1080 nm is 10 mW. When NOPAs operate at deamplification (the pump
light and the injected signals are $\pi $ out of phase), the intensity of
each submode of the obtained cluster state is about $30$ $\mu $W$.$ The
power of the local oscillation beam in each HD is about $4$ mW and the
intensity of LO1 (LO2) is $54$ $\mu $W. All the squeezed correlation
variances of the cluster state measured under the previous conditions are
about 3 dB below the corresponding SNL (the equivalent squeezing parameter $r
$\ equals to $0.35$). The electronic gain $g$\ in classical channels is
carefully adjusted to $1.00\pm 0.05$\ according to the method described in
Ref. \cite{Zhang2003}.

\begin{figure}[tbp]
\begin{center}
\includegraphics[width=7cm,height=6cm]{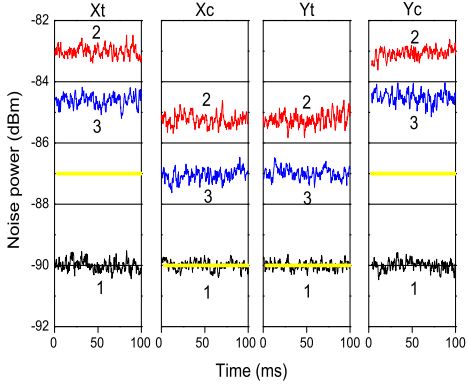}
\end{center}
\caption{(Color online) Noise power of the quadratures with vacuum
inputs. Black (1) lines: noise of vacuum state; red (2) lines: noise
variances of output quadratures without cluster state; blue (3)
lines: noise variances of output quadratures with cluster state; and
yellow (straight) lines: noise variances of output quadratures in
the ideal case.}
\end{figure}

\section{Experimental results}

To quantify the performance of the controlled-X operation, the noise
variances of $\hat{X}_{t(c)}^{out}$ and $\hat{Y}_{t(c)}^{out}$ in Eqs. $%
(6)-(9)$ are measured by HD5(6) and are recorded by a following spectrum
analyzer (SA) with the resolution bandwidth of $30$ kHz and video bandwidth
of $100$ Hz. The measured noise powers of $\hat{X}_{t(c)}^{out}$ and $\hat{Y}%
_{t(c)}^{out}$ with the vacuum input ($\hat{a}_{t}$ and $\hat{a}_{c}$ are
vacuum state) are shown in Fig. 2. The noise powers of the two input vacuum
states serve as the SNL [(black) 1 lines]. In case of infinite squeezing
[(yellow) straight lines], the noise variances of $\hat{X}_{t}^{out}$\ and $%
\hat{Y}_{c}^{out}$\ are 3 dB above the SNL due to the effect of $\hat{X}_{c}$
($\hat{Y}_{t}$), while noise variances of $\hat{X}_{c}^{out}$\ and $\hat{Y}%
_{t}^{out}$ remain at SNL [see Eq. (1)]. The (red) 2 lines and the (blue) 3
lines illustrate the performance of the controlled-X operation without and
with using the quantum entanglement of the cluster state, respectively. The
(red) 2 lines are measured by replacing each cluster submode with a coherent
light of identical intensity. The measured\textbf{\ }values of $\hat{X}%
_{t}^{out}$, $\hat{X}_{c}^{out}$, $\hat{Y}_{t}^{out}$, and $\hat{Y}_{c}^{out}
$\ are 6.95, 4.76, 4.77 and 6.93 dB above the SNL, respectively. The
variances of $\hat{X}_{t}^{out}$, $\hat{X}_{c}^{out}$, $\hat{Y}_{t}^{out}$,
and $\hat{Y}_{c}^{out}$ measured with the existence of the cluster state
[(blue) 3 lines] are 5.39, 2.95, 3.01, 5.50 dB above the SNL, respectively.
The (blue) 3 lines with the cluster state are $\sim 1.8$ dB below that
without using the cluster [(red) 2 lines]. It means that the precision of
the controlled-X operator increases about $1.8$ dB with respect to its
classical copies. We use the fidelity formula $F=\left\{ \text{Tr}[(\sqrt{%
\hat{\rho}_{1}}\hat{\rho}_{2}\sqrt{\hat{\rho}_{1}})^{1/2}]\right\} ^{2}$,
which denotes the overlap between the experimental obtained output state $%
\hat{\rho}_{2}$\ and ideal output sate $\hat{\rho}_{1}$, to quantify the
performance of controlled-X operation. The fidelity for two Gaussian states $%
\hat{\rho}_{1}$\ and $\hat{\rho}_{2}$\ with covariance matrices $\mathbf{A}%
_{i}$\ and mean amplitudes $\mathbf{\alpha }_{i}\equiv (\alpha _{iX},\alpha
_{iY})$\ ($i$=$1,2$)\ is expressed by \cite{Nha2005,Scutaru1998}
\begin{equation}
F=\frac{2}{\sqrt{\Delta +\delta }-\sqrt{\delta }}\exp [-\mathbf{\beta }^{T}(%
\mathbf{A}_{1}+\mathbf{A}_{2})^{-1}\mathbf{\beta }],
\end{equation}%
\ where $\Delta =\det (\mathbf{A}_{1}+\mathbf{A}_{2}),$\ $\delta =(\det
\mathbf{A}_{1}-1)(\det \mathbf{A}_{2}-1),$\ $\mathbf{\beta }=\mathbf{\alpha }%
_{2}-\mathbf{\alpha }_{1},$ $\mathbf{A}_{1}$ and $\mathbf{A}_{2}$ for the
ideal ($\hat{\rho}_{1}$) and experimental ($\hat{\rho}_{2}$) output states,
respectively. For our case, the covariance matrices $\mathbf{A}_{i}$\ ($i$=$%
1,2$) for the target mode are given by

\begin{eqnarray}
\mathbf{A}_{1} &=&\left[
\begin{array}{cc}
V(X_{t}-X_{c}) & 0 \\
0 & V(Y_{t})%
\end{array}%
\right] =\left[
\begin{array}{cc}
2 & 0 \\
0 & 1%
\end{array}%
\right] ,\qquad  \\
\mathbf{A}_{2} &=&\left[
\begin{array}{cc}
V(X_{t}^{out}) & 0 \\
0 & V(Y_{t}^{out})%
\end{array}%
\right] .
\end{eqnarray}%
Similarly, we can write out the covariance matrices for the control mode.$\ $%
In case of infinite squeezing, both fidelities for the control and target
states $F_{c}$ and $F_{t}$ equal 1, which can be calculated from Eqs. $%
(6)-(9)$\ with $r\rightarrow \infty $. Without using cluster resources ($r=0$%
), the fidelity for both $\hat{a}_{t}$ and $\hat{a}_{c}$ states is 0.73.
When the cluster states are used, the obtained fidelities of $\hat{a}_{t}$
and $\hat{a}_{c}$ are both 0.87, which are about 0.14 better than those
without using entanglement.

\begin{figure}[tbp]
\begin{center}
\includegraphics[width=4cm,height=4cm]{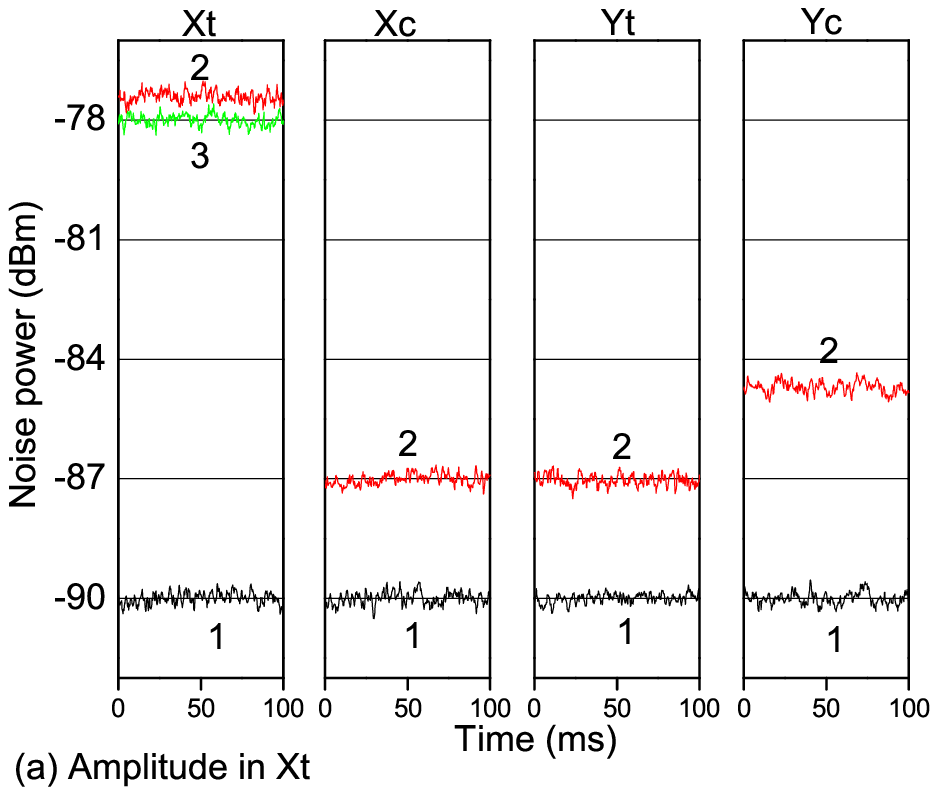} %
\includegraphics[width=4cm,height=4cm]{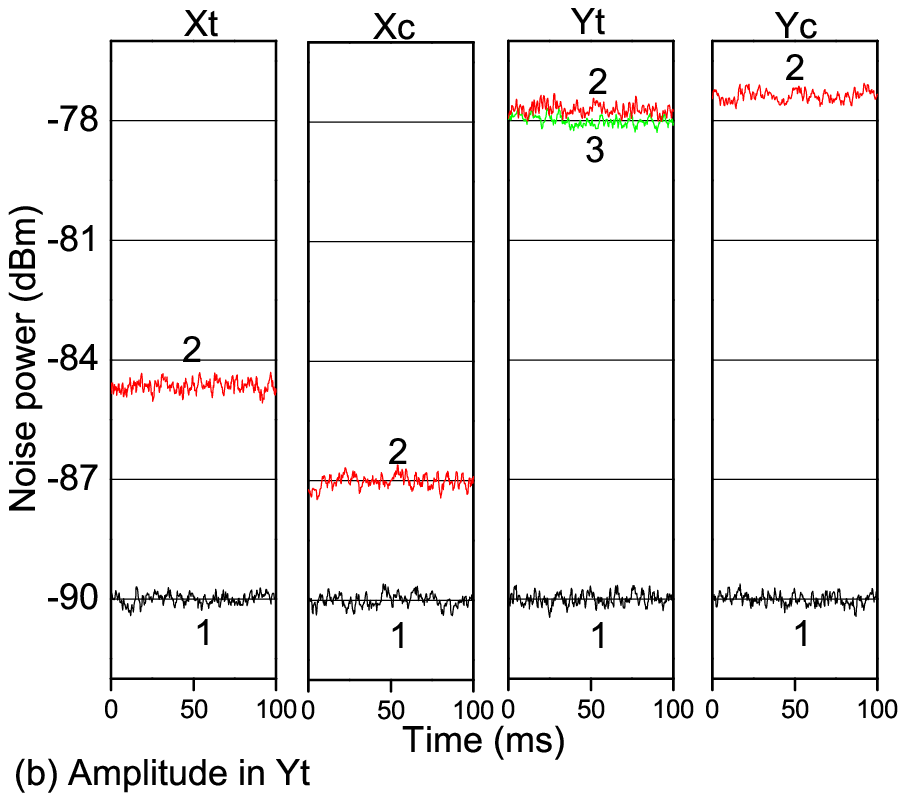} %
\includegraphics[width=4cm,height=4cm]{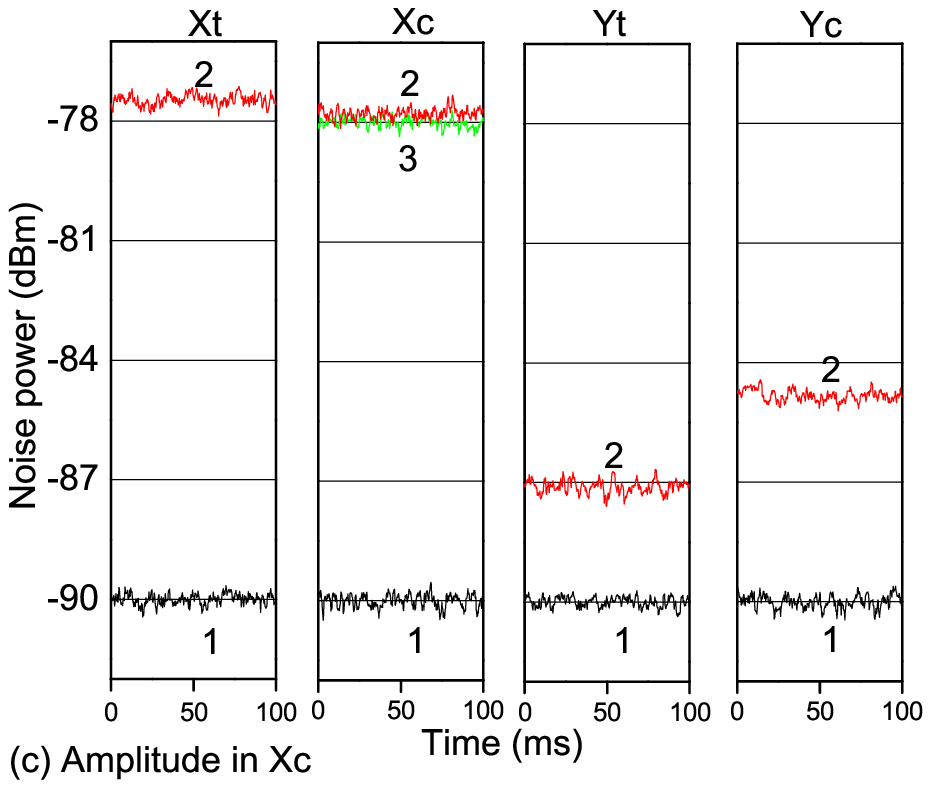} %
\includegraphics[width=4cm,height=4cm]{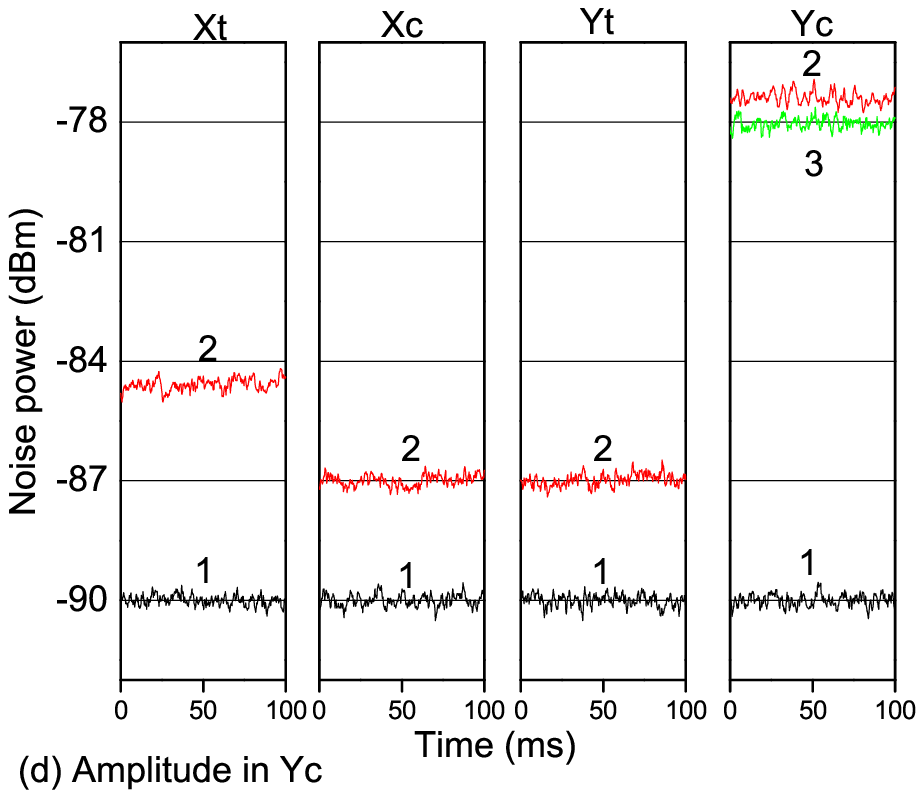}
\end{center}
\caption{(Color online) Noise power of the quadratures with four
different coherent input quadratures. Black (1) lines: noises of
vacuum state; red (2) lines: output variances with cluster state;
and green (3) lines: input variances.}
\end{figure}

To simulate the performance of the controlled-X gate under the coherent
input signals with nonzero average intensity, we modulate the input $\hat{a}%
_{t}$ and $\hat{a}_{c}$ optical beams with amplitude and phase modulators at
$2$ MHz, (not shown in Fig. 1). The coherent excitations of four different
quadratures corresponding to four input states with individual modulation
are investigated. The traces in Figs. 3 (a), (b), (c) and (d) correspond to
four different input states: (a) and (b) for the amplitude and phase
modulation of the input target $a_{t}$ and (c) and (d) for the amplitude and
phase modulation of the input control $a_{c}$, respectively. In the four
different modulation models, the intensity of the imposed modulation signals
is identical. The (black) 1 lines and the (green) 3 lines in Fig. 3
correspond to the SNL and the quadratures of the input signals,
respectively. The quadratures of the outcome fields are shown with the (red)
2 lines. All coherent amplitudes of the input quadratures are $12$ dB above
the SNL due to the same modulation strength, which are measured by HD5 and
HD6 in the case blocking the cluster beams. We can see from Figs. 3 (b) and
(c) that the quadrature $\hat{Y}_{t}$ $(\hat{X}_{c})$ of the input state is
coupled into $\hat{Y}_{c}^{out}$ ($\hat{X}_{t}^{out}$) of the outcome state,
while $\hat{Y}_{c}=\hat{Y}_{c}^{out}$ ($\hat{X}_{t}=\hat{X}_{t}^{out}$) is
preserved. Figs. 3 (a) and (d) show that the quadratures $\hat{X}_{t}$ and $%
\hat{Y}_{c}$ of input states are not coupled to any output quadratures, in
which, however, the variances of the outcome quadratures are a little higher
than that of the input states due to the effect of imperfect squeezing.

\section{Conclusion}

In conclusion, an experimental study toward demonstrating the CV
controlled-X logical operation based on four-partite cluster states of
electro-magnetic fields is presented. In the experiment the information
encoded in the input target and control states are teleported to the outcome
states via quantum channels depending on multipartite cluster entanglement
among quadrature components of optical modes and electronic classical
channels. The one-way controlled-X gate can be regarded as to be constructed
by two CV quantum teleporters based on CV quantum entanglement among modes $%
\hat{b}_{1}$, $\hat{b}_{2\text{, }}\hat{b}_{3}$ and $\hat{b}_{4}$, $\hat{b}%
_{2\text{, }}\hat{b}_{3},$ respectively. The difference between these
teleporters and the normal teleportation systems \cite%
{Furusawa1998,Bowen2003,Zhang2003} is that there are two input states
(target and control) in these teleporters and both of their information is
transmitted to the two output states simultaneously. To achieve the basic
operation of one-way quantum computation the four-partite cluster
entanglement plays irreplaceable roles. In CV QCs, we have to establish
entanglement between the output target and control modes for further quantum
information processing. According to the theoretical calculation \cite%
{Duan2000} for achieving the entanglement of the two output states the
initial squeezing degree of cluster resources should be higher than $\sim $
7 dB. Although the entanglement between output states was not observed in
the present experiment due to the absence of better cluster resources, the
proof-of-principle experiment proves that the controlled-X operation of
one-way QC can be unconditionally demonstrated with the designed system if
CV cluster states of higher entanglement are available.

This research was supported by the NSFC (Grant Nos. 60736040 and 10804065),
NSFC Project for Excellent Research Team (Grant No. 60821004), and the
National Basic Research Program of China (Grant No. 2007BAQ03918).


\begin{thebibliography}{99}
\bibitem{Monroe1995} C. Monroe, D. M. Meekhof, B. E. King, W. M. Itano, and
D. J. Wineland, Phys. Rev. Lett. \textbf{75}, 4714 (1995).

\bibitem{Pittman2002} T. B. Pittman, B. C. Jacobs, and J. D. Franson, Phys.
Rev. Lett. \textbf{88}, 257902 (2002).

\bibitem{Kaler2003} F. Schmidt-Kaler, H. H\"{a}ffner, M. Riebe, S. Gulde,
G. P. T. Lancaster, T. Deuschle, C. Becher, C. F. Roos, J. Eschner,
and R. Blatt, Nature, \textbf{422}, 408, (2003).

\bibitem{Leibfried2003} D. Leibfried, B. DeMarco, V. Meyer, D. Lucas, M.
Barrett, J. Britton, W. M. Itano, B. Jelenkovic, C. Langer, T. Rosenband,
and D. J. Wineland, Nature, \textbf{422}, 412, (2003).

\bibitem{Li2003} X. Li, Y. Wu, D. Steel, D. Gammon, T. H. Stievater, D. S.
Katzer, D. Park, C. Piermarocchi, and L. J. Sham, Science, \textbf{301},
809, (2003).

\bibitem{Zhao2005} Z. Zhao, A. N. Zhang, Y. A. Chen, H. Zhang, J. F. Du, T.
Yang, and J. W. Pan, Phys. Rev. Lett. \textbf{94}, 030501 (2005).

\bibitem{Yoshikawa2008} J. I. Yoshikawa, Y. Miwa, A. Huck, U.~L. Andersen,
P. van Loock, and A. Furusawa, Phys. Rev. Lett. \textbf{101}, 250501 (2008).

\bibitem{Miwa2009} Y. Miwa, J. I. Yoshikawa, P. van Loock, and A. Furusawa,
Phys. Rev. A \textbf{80}, 050303(R), (2009).

\bibitem{Nielsen2000} M. A. Nielsen and I. L. Chuang, \textit{Quantum
computation and Quantum information} (Cambridge University Press, Cambridge,
2000).

\bibitem{Chuang1998} I. L. Chuang, L. M. K. Vandersypen, X. Zhou, D. W.
Leung, S. Lloyd, Nature, \textbf{393}, 143, (1998).

\bibitem{Gottesman1999} D. Gottesman, and I. L. Chuang, Nature, \textbf{402}%
, 390, (1999).

\bibitem{Gulde2003} S. Gulde, M. Riebe, G. P. T. Lancaster, C. Becher, J.
Eschner, H. H\"{a}ffner, F. Schmidt-Kaler, I. L. Chuang, and R. Blatt,
Nature, \textbf{421}, 48, (2003).

\bibitem{Lu2007} C. Y. Lu, D. E. Browne, T. Yang, and J. W. Pan, Phys. Rev.
Lett. \textbf{99}, 250504 (2007).

\bibitem{Politi2009} A. Politi, J. C. F. Matthews, J. L. O'Brien, Science,
\textbf{325}, 1221 (2009).

\bibitem{Raussendorf2001} R. Raussendorf and H.~J. Briegel, Phys. Rev. Lett.
\textbf{86}, 5188 (2001).

\bibitem{Walther2005} P. Walther, K. J. Resch, T. Rudolph, E. Schenck, H.
Weinfurter, V. Vedral, M. Aspelmeyer, and A. Zeilinger, Nature, \textbf{434}%
, 169 (2005).

\bibitem{Prevedel2007} R. Prevedel, P. Walther, F. Tiefenbacher, P. B\"{o}%
hi, R. Kaltenbaek, T. Jennewein and A. Zeilinger, Nature, \textbf{445}, 65
(2007).

\bibitem{Chen2007} K. Chen, C. M. Li, Q. Zhang, Y. A. Chen, A. Goebel, S.
Chen, A. Mair, and J. W. Pan, Phys. Rev. Lett. \textbf{99}, 120503 (2007).

\bibitem{Lloyd1999} S. Lloyd and S.~L. Braunstein, Phys. Rev. Lett. \textbf{%
82}, 1784 (1999).

\bibitem{Ralph2003} T.~C. Ralph, A. Gilchrist, G.~J. Milburn, W.~J. Munro,
and S. Glancy, Phys. Rev. A \textbf{68}, 042319 (2003).

\bibitem{Gottesman2001} D. Gottesman, A. Kitaev, and J. Preskill, Phys. Rev.
A \textbf{64}, 012310 (2001).

\bibitem{Braunstein1998} S. L. Braunstein, Phys. Rev. Lett. \textbf{80},
4084 (1998).

\bibitem{Aoki2009} T. Aoki, G. Takahashi, T. Kajiya, J. Yoshikawa, S. L.
Braunstein, P. van Loock, A. Furusawa, Nature Physics \textbf{5}, 541 (2009).

\bibitem{Filip2005} R. Filip, P. Marek, and U. L. Andersen, Phys. Rev. A
\textbf{71}, 042308 (2005).

\bibitem{Zhang2006} J. Zhang and S.~L. Braunstein, Phys. Rev. A \textbf{73},
032318 (2006).

\bibitem{vanLoock2007} P. van Loock, C. Weedbrook, and M. Gu, Phys. Rev. A
\textbf{76}, 032321 (2007).

\bibitem{Menicucci2006} N.~C. Menicucci, P. {van Loock}, M. Gu, C.
Weedbrook, T.~C. Ralph, and M.~A. Nielsen, Phys. Rev. Lett. \textbf{97},
110501 (2006).

\bibitem{vanLoockJOSA2007} P. van Loock, J. Opt. Soc. Am. B \textbf{24}, 340
(2007).

\bibitem{Tan2009} A. H. Tan, C. D. Xie, and K. C. Peng, Phys. Rev. A \textbf{%
79}, 042338 (2009).

\bibitem{Gu2009} M. Gu, C. Weedbrook, N. C. Menicucci, T. C. Ralph, and P.
van Loock, Phys. Rev. A \textbf{79}, 062318 (2009).

\bibitem{Su2007} X. L. Su, A. H. Tan, X. J. Jia, J. Zhang, C. D. Xie, and K.
C. Peng, Phys.\ Rev.\ Lett. \textbf{98}, 070502 (2007).

\bibitem{Yukawa2008} M. Yukawa, R. Ukai, P. van Loock, and A. Furusawa,
Phys. Rev. A \textbf{78}, 012301 (2008).

\bibitem{Lloyd1992} S. Lloyd, Phys. Let. A \textbf{167}, 255 (1992).

\bibitem{Hemmer2009} P. Hemmer and J. Wrachtrup, Science, \textbf{324}, 473 (2009).

\bibitem{Bouwmeester1997} D. Bouwmeester, J. W. Pan, K. Mattle, M. Eibl, H.
Weinfurter, and A. Zeilinger, Nature, \textbf{390}, 575 (1997).

\bibitem{Boschi1998} D. Boschi, S. Branca, F. De Martini, L. Hardy, and S.
Popescu, Phys. Rev. Lett., \textbf{80}, 1121 (1998).

\bibitem{Riebe2004} M. Riebe, H. H\"{a}ffner, C. F. Roos, W. H\"{a}nsel, J.
Benhelm, G. P. T. Lancaster, T. W. K\"{o}rber, C. Becher, F. Schmidt-Kaler,
D. F. V. James, and R. Blatt, Nature, \textbf{429}, 734 (2004).

\bibitem{Barrett2004} M. D. Barrett, J. Chiaverini, T. Schaetz, J. Britton,
W. M. Itano, J. D. Jost, E. Knill, C. Langer, D. Leibfried, R. Ozeri, and D.
J. Wineland, Nature, \textbf{429}, 737 (2004).

\bibitem{Huang2004} Y. F. Huang, X. F. Ren, Y. S. Zhang, L. M. Duan, and G.
C. Guo, Phys. Rev. Lett, \textbf{93}, 240501 (2004).

\bibitem{Sherson2006} J. F. Sherson, H. Krauter, R. K. Olsson, B. Julsgaard,
K. Hammerer, I. Cirac, and E. S. Polzik, Nature, \textbf{443}, 557 (2006).

\bibitem{Olmschenk2009} S. Olmschenk, D. N. Matsukevich, P. Maunz, D. Hayes,
L.-M. Duan, and C. Monroe, Science, \textbf{323}, 486 (2009).

\bibitem{Furusawa1998} A. Furusawa, J. L. Sorenson, S. L. Braunstein, C. A.
Fuchs, H. J. Kimble, E. S. Polzik, Science, \textbf{282}, 706 (1998).

\bibitem{Bowen2003} W. P. Bowen, N. Treps, B. C. Buchler, R. Schnabel, T. C.
Ralph, Hans-A. Bachor, T. Symul, and P. K. Lam, Phys. Rev. A \textbf{67},
032302 (2003).

\bibitem{Zhang2003} T. C. Zhang, K. W. Goh, C. W. Chou, P. Lodahl, and H. J.
Kimble, Phys. Rev. A, \textbf{67}, 033802 (2003).

\bibitem{Zhao2004} Z. Zhao, Y. A. Chen, A. N. Zhang, T. Yang, H. J. Briegel,
and J. W. Pan, Nature, \textbf{430}, 54 (2004).

\bibitem{Yonezawa2004} H. Yonezawa, T. Aoki, and A. Furusawa, Nature,
\textbf{431}, 430 (2004).

\bibitem{Li2002} X. Y. Li, Q. Pan, J. T. Jing, J. Zhang, C. D. Xie, and K.
C. Peng, Phys.\ Rev.\ Lett. \textbf{88}, 047904 (2002).

\bibitem{Jia2004} X. J. Jia, X. L. Su, Q. Pan, J. R. Gao, C. D. Xie, and K.
C. Peng, Phys.\ Rev.\ Lett. \textbf{93}, 250503 (2004).

\bibitem{Nha2005} H. Nha, and H. J. Carmichael, Phys. Rev. A, \textbf{71},
032336 (2005).

\bibitem{Scutaru1998} H. Scutaru, J. Phys. A \textbf{31}, 3659 (1998).

\bibitem{Duan2000} Lu-Ming Duan, G. Giedke, J. I. Cirac, and P. Zoller,
Phys. Rev. Lett. \textbf{84}, 2722 (2000).
\end{thebibliography}
\end{document}